\documentclass[twocolumn,superscriptaddress,aps,showpacs,floatfix,pra,10pt]{revtex4-1}
\usepackage{amssymb,amsmath}
\usepackage{hyperref,graphicx}
\usepackage{dcolumn}
\usepackage{color}
\usepackage{ulem}
\usepackage{bbold}
\newcommand{\bs}[1]{{\boldsymbol{#1}}}
\newcommand{\bk}{\bs{k}}
\newcommand{\bq}{\bs{q}}
\newcommand{\av}[1]{\overline{#1}}
\newcommand{\rmd}{\mathrm{d}}

\newcommand{\intdtwo}[1]{\int \frac{\rmd^2 #1}{(2\pi)^2}}
\newcommand{\tauh}{\tau_H}
\newcommand{\taub}{\tau_B}
\newcommand{\taud}{\tau_D}
\newcommand{\intdone}[1]{\int \frac{\rmd #1}{2\pi}}

\begin{document}

\title{Coherent forward scattering in 2D disordered systems}

\author{S. Ghosh}
\affiliation{Centre for Quantum Technologies, National University of Singapore, 3 Science Drive 2, Singapore 117543, Singapore}
\affiliation{Laboratoire Kastler Brossel, UPMC Univ Paris 06, Ecole Normale Sup\'{e}rieure, CNRS, Coll\`{e}ge de France; 4 Place Jussieu, 75005 Paris, France}
\affiliation{Merlion MajuLab, CNRS-UNS-NUS-NTU International Joint Research Unit, UMI 3654, Singapore}
\author{N. Cherroret}
\email[Corresponding author: ]{cherroret@lkb.upmc.fr}
\affiliation{Laboratoire Kastler Brossel, UPMC Univ Paris 06, Ecole Normale Sup\'{e}rieure, CNRS, Coll\`{e}ge de France; 4 Place Jussieu, 75005 Paris, France}
\author{B. Gr\'{e}maud}
\affiliation{Merlion MajuLab, CNRS-UNS-NUS-NTU International Joint Research Unit, UMI 3654, Singapore}
\affiliation{Centre for Quantum Technologies, National University of Singapore, 3 Science Drive 2, Singapore 117543, Singapore}
\affiliation{Department of Physics, National University of Singapore, 2 Science Drive 3, Singapore 117542, Singapore}
\author{C. Miniatura}
\affiliation{Merlion MajuLab, CNRS-UNS-NUS-NTU International Joint Research Unit, UMI 3654, Singapore}
\affiliation{Centre for Quantum Technologies, National University of Singapore, 3 Science Drive 2, Singapore 117543, Singapore}
\affiliation{Department of Physics, National University of Singapore, 2 Science Drive 3, Singapore 117542, Singapore}
\affiliation{INLN, Universit\'{e} de Nice-Sophia Antipolis, CNRS; 1361 route des Lucioles, 06560 Valbonne, France}
\author{D. Delande}
\affiliation{Laboratoire Kastler Brossel, UPMC Univ Paris 06, Ecole Normale Sup\'{e}rieure, CNRS, Coll\`{e}ge de France; 4 Place Jussieu, 75005 Paris, France}

\begin{abstract}
We present a detailed numerical and theoretical analysis of the recently discovered phenomenon of coherent forward scattering. This effect manifests itself as a macroscopic interference peak in the forward direction of the momentum distribution of a matter wave launched with finite velocity in a random potential. Focusing on the two-dimensional case, we show that coherent forward scattering generally arises due the confinement of the wave in a finite region of space, and explain under which conditions it can be seen as a genuine signature of Anderson localization.

\end{abstract}



\maketitle

\section{Introduction}

In the last fifty years, the physics of disordered systems has turned out to be tremendously rich, and the field is still offering challenging and unexpected results. Among those, the manifestations of weak and strong (Anderson) localization of coherent waves are paradigmatic examples \cite{Anderson58, Bergmann84}. In practice, Anderson localization very often manifests itself as a halt of wave transport. This signature has been largely exploited in a number of experiments searching for Anderson localization of classical waves in disordered media \cite{Schwartz07, Hu08, Maret12} or of matter waves subjected to time-periodic \cite{Chabe08, Lemarie10} and random \cite{Billy08, Roati08, Kondov11, Jendrzejewski12, Semeghini14} optical potentials. At the same time, recent works \cite{Cherroret12, Karpiuk12, Micklitz14, Loon14} pointed out that in ultracold-atom setups, the momentum distribution of a matter wave in a random potential can exhibit a highly nontrivial dynamics due to localization if it is initially 
launched with a nonzero mean wave-vector $\bk_0$. The scenario is then the following. First, over a time scale of the order of the Boltzmann transport mean free time $\tau_B$, an isotropization of the distribution takes place as particles' momenta are being randomized by the disorder \cite{Plisson13}. During this process, a narrow coherent backscattering (CBS) peak emerges around the direction $-\bk_0$. After a few $\tau_B$, this peak gains a maximum visibility and sits on top of a broader isotropic, ring-shaped distribution \cite{Cherroret12}. At a later time, a second interference peak appears in the forward direction $+\bk_0$ \cite{Karpiuk12}. The visibility of this coherent forward scattering (CFS) peak  increases slowly, and finally reaches a maximum value at a time of the order of the Heisenberg time $\tau_H$, defined as the inverse of the mean spacing between the energy levels of the system 
(see below for a more precise discussion). Beyond $\tau_H$, the system no longer evolves and the asymptotic distribution has the central 
symmetry (if time-reversal invariance is preserved), with two identical CBS and CFS peaks. Experimentally, the CBS effect of ultracold atoms has been recently observed \cite{Josse12}, motivating further investigations like the sensitivity of CBS to external dephasing \cite{Cord14}. An observation of the CFS effect is, on the other hand, still missing. From a theoretical point of view, while the physics of CBS is today well understood -- it stems from wave amplitudes travelling along the same multiple scattering sequence but in opposite directions \cite{Aegerter09} --
the mechanism of CFS is much less obvious. In fact, the building of the CFS peak relies on interference sequences where particles are scattered back and forth several times between the directions $-\bk_0$ and $+\bk_0$ \cite{Karpiuk12, Micklitz14, Loon14}. While this mechanism is inefficient in a purely diffusive system, it becomes strongly enhanced when the wave gets confined in a limited region of space: wave interference can then accumulate and make the CFS peak macroscopic. The situation typically occurs in an infinite system if Anderson localization comes into play (this was the scenario originally considered in \cite{Karpiuk12}), but also, as discussed in the present paper, when the wave is trapped within a limited region of space of size smaller than the localization length. In this respect, a full characterization of the CFS effect is required in order to unambiguously attribute its appearance to Anderson localization.

In this paper, we present a detailed study of the momentum distribution of a matter wave launched in a random potential, bringing special attention to the CFS effect to which we propose a systematic numerical analysis combined with theoretical predictions. We consider a random potential of speckle type, routinely used in current experiments with ultracold atoms \cite{ColdDisorderRevs}, and focus on the two-dimensional (2D) geometry for which the CFS peak clearly distinguishes itself from the isotropic part of the momentum distribution \cite{Karpiuk12} (we refer the reader to \cite{Loon14, Micklitz14} for the one-dimensional and quasi one-dimensional cases). The main concepts discussed in the paper as well as the theoretical framework are introduced in Sec. \ref{framework}. In Sec. \ref{sectionL}, we analyze the building up of the CFS peak for a matter wave scattered diffusively in a limited volume of size much smaller than the localization length of the problem, $L\ll\xi$. Then, in Sec. \ref{sectionxi}, we address the more difficult but more interesting case $\xi\ll L$ where the CFS peak is triggered by the confinement of the matter wave stemming from Anderson localization. This configuration is typically the one of experiments, where an atomic wave packet is initially prepared and evolves in the presence of the disorder, without any confining box. This experimental scenario, studied in Sec.~\ref{feasibility}, has however two additional ingredients: (i) the matter wave has a broad energy distribution for which both localized and diffusive atoms coexist and (ii) the initial state has a finite spatial width, affecting the shape and height of the CFS peak. Despite these complications, we show that, in the absence of an artificial confining box, the diffusive components studied in Sec.~\ref{sectionL} do not contribute to the CFS signal, and that the latter can be observed and used as a ``smoking gun'' of Anderson localization. 
In Sec. \ref{Conclusion}, we finally summarize our results and discuss some open questions.

\section{Momentum distribution of a matter wave in a 2D random potential}
\label{framework}

\subsection{Numerical experiment}
\label{numerics}

In order to introduce the physics discussed in the paper, let us first consider a simple numerical experiment. We start from a plane matter wave $|\bk_0\rangle$ and propagate it with the evolution operator $\exp(-i\hat{H}t)$, where $\hat{H}=\bs{p}^2/(2m)+V(\textbf{r})$ with $V(\textbf{r})$ a 2D random potential (from here on we set $\hbar=1$). Following recent experiments on ultracold atoms, we choose $V(\textbf{r})$ to be a blue-detuned speckle potential with mean value $\av{V(\textbf{r})}=0$ and correlation function $\av{V(\textbf{r})V(\textbf{r}')}=[2V_0J_1(|\textbf{r}-\textbf{r}'|/\zeta)/(|\textbf{r}-\textbf{r}'|/\zeta)]^2$, with $\zeta$ the correlation length. This potential is numerically generated in a standard way, by convoluting a circular Gaussian random field with a cutoff function  that simulates the diffusive plate used in experiments \cite{Huntley89, Horak98}. As soon as the random potential is turned on, $\bk_0$ is no longer a good quantum number and the system starts 
to evolve. The time propagation is achieved on a 2D grid of size $L\times L$ with periodic boundary conditions along $x$ and $y$, by using an iterative method based on the expansion of the evolution operator in combinations of Chebyshev polynomials of the Hamiltonian \cite{Fehske09, Roche97}. In the simulations, a cell of surface $(\pi\zeta)^2$ is discretized in typically 8-10 steps along both $x$ and $y$. After the evolution, the momentum distribution is calculated by applying a discrete Fourier transformation on the final wave function. This procedure is repeated for many disorder realizations, which finally gives access to the disorder-averaged momentum distribution. Throughout the paper, lengths, momenta, energies and times will be given in units of $\zeta$, $\zeta^{-1}$, $1/(m\zeta^2)$ and $m\zeta^2$, respectively. 

\begin{figure}
\includegraphics[width=0.72\linewidth]{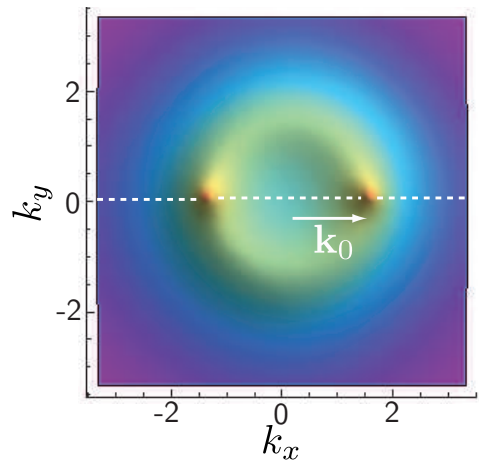}
\caption{(Color online) Density plot of the long-time momentum distribution obtained after numerical propagation of a plane wave $|\bk_0\rangle$ in a 2D speckle potential. Parameters are $k_0=1.5$, $V_0=5$, $L=20\pi$ and $t=10^3$, where lengths, momenta, energies and times are given in units of $\zeta$, $\zeta^{-1}$, $1/(m\zeta^2)$ and $m\zeta^2$, respectively. The left peak is due to coherent backscattering and the right peak to coherent forward scattering. Data are averaged over 7200 disorder realizations.
} 
\label{twinpeaks}
\end{figure}

A typical distribution obtained at long times (here $t=10^3$) is shown in Fig. \ref{twinpeaks}, for $V_0=5$, $k_0\equiv|\bk_0|=1.5$ and for a system size $L=20\pi$. 
The distribution of Fig. \ref{twinpeaks}, $\overline{n}(\bk)=\overline{n}_\text{D}+\overline{n}_\text{CBS}+\overline{n}_\text{CFS}$, exhibits three components: an isotropic, diffusive ring-shaped ``background'' $\overline{n}_\text{D}$ and two interference peaks $\overline{n}_\text{CFS}$ and $\overline{n}_\text{CBS}$ centered at $\pm \bk_0$. The ring describes the quasi-elastic isotropization of atomic momenta in the course of the propagation in the random potential. The peak centered at $-\bk_0$ is the coherent backscattering peak, and the peak centered at $+\bk_0$ is the coherent forward scattering peak \cite{Cherroret12, Karpiuk12}. 

The stationary distribution of Fig. \ref{twinpeaks} is obtained after a long time of propagation in the random potential. Before this final state establishes however, the system explores a number of regimes summarized in Fig. \ref{phase_diagram}, and that we discuss below.

\subsection{Diagrammatic theory in the diffusive regime}
\label{theory_diag}

At short times, a diagrammatic description of the momentum distribution can be developed. We briefly summarize below the essential steps of this approach. In momentum space, the wave function $|\psi(t)\rangle$ at time $t$ reads
\begin{equation}
\langle\bk|\psi(t)\rangle=\int \frac{d^2\bk' dE_1}{(2\pi)^3} e^{-iE_1 t} \langle\bk|\hat{G}^R(E_1)|\bk'\rangle
 \langle\bk'|\phi\rangle,
\end{equation}
where $|\phi\rangle$ is the wave function at time $t=0$ and $\hat{G}^R(E_1)=(E_1-\hat{H}+i0^+)^{-1}$ is the retarded Green's operator at energy $E_1$. The disorder-averaged momentum distribution at time $t$, $\av{n}(\bk,t)=\av{|\langle\bk|\psi(t)\rangle|^2}$, then involves the disorder-averaged intensity propagation kernel $\av{\langle\bk |\hat{G}^R(E_1)|\bk'\rangle\langle\bk''| \hat{G}^A(E_2)|\bk'\rangle}$, integrated over the momenta $\bk'$ and $\bk''$ and over the energies $E_1$ and $E_2$ [$\hat{G}^A(E_1)=(E_1-\hat{H}-i0^+)^{-1}$ is the advanced Green's operator]. Since the disorder average restores translation invariance and thus momentum conservation, this kernel takes the simpler form $(2\pi)^2\delta(\bk'-\bk'')\Phi_{\bk'\bk E}(\omega)$, where $E=(E_1+E_2)/2$, $\omega=E_1-E_2$, and where $\Phi_{\bk'\bk E}$ remains to be determined. The disorder-averaged momentum distribution at time $t$ thus reads \cite{Kuhn07}
\begin{equation}
\av{n}(\bk,t) = \intdone{\omega}e^{-i\omega t} \int \frac{d^2\bk' dE}{(2\pi)^3} \Phi_{\bk'\bk E}(\omega)n_0(\bk'),
\end{equation}
where $n_0(\bk')=| \langle\bk'|\phi\rangle|^2$.
In the numerical simulation, we consider for simplicity an initial plane wave $|\phi\rangle = |\bk_0\rangle$, leading to $n_0(\bk')=(2\pi)^2\delta(\bk'-\bk_0)$ and to
\begin{equation} \label{rhokprimet}
\av{n}(\bk,t) = \intdone{\omega} e^{-i\omega t}\intdone{E}\Phi_{\bk_0\bk E}(\omega).
\end{equation}

In this section and the next one, we discuss the dynamics at a given energy $E$. To lighten the notations we therefore drop the $E-$dependence of all time and length scales, keeping in mind that the full momentum distribution is a superposition of many energy components [see Eq. (\ref{rhokprimet})], each of which behaving \emph{a priori} differently in the disorder. We also assume that disorder is weak, namely $k_0\ell_B\gg 1$, where  $\ell_B$ is the Boltzmann transport mean free path \cite{AM}. In this limit, the various time scales of the system are well separated, and several regimes of transport can be clearly identified, as we now discuss.
\begin{figure}[h]
\includegraphics[width=0.75\linewidth]{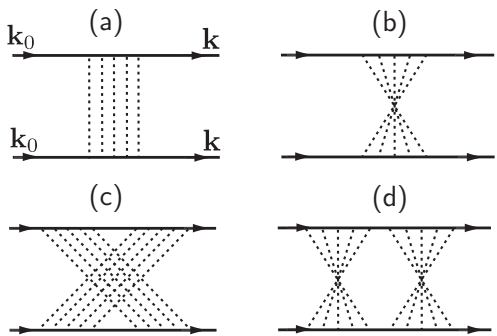}
\caption{Leading-order diagrams contributing to the momentum distribution. (a): series of ladder diagrams giving rise to the isotropic part of the distribution, Eq. (\ref{Background}). (b): series of crossed diagrams giving rise to the CBS peak, Eq. (\ref{CBS_diag}). The two series of diagrams (c) and (d) give the main contribution to the CFS peak in the diffusive regime $\taub\ll t\ll \text{min}(L,\xi)^2/D_B$, Eq. (\ref{CFS_diag}). They equally contribute for a time-reversal invariant system. } 
\label{diagrams}
\end{figure}

\begin{figure}
\includegraphics[width=0.75\linewidth]{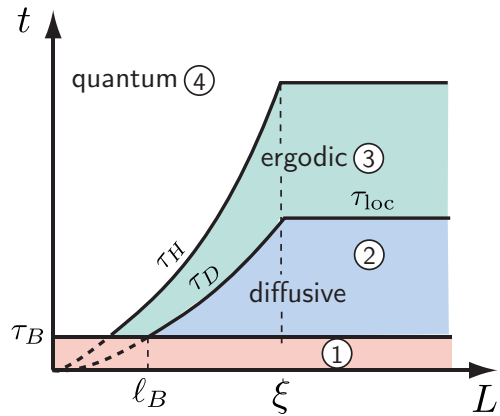}
\caption{(Color online) Schematic dynamical phase diagram of a 2D, weakly disordered system. $\taub$ is the transport mean free time and $\ell_B$ the transport mean free path. For $t\lesssim \taub$ atoms undergo a few scattering events and the distribution gets isotropized (regime 1). For $\taub\ll t\ll \taud, \tau_\text{loc}$, the randomization process is completed and transport is diffusive (regime 2). Beyond the Thouless time $\taud\equiv L^2/D_B$ (for $L<\xi$) or the localization time $\tau_\text{loc}\equiv  \xi^2/D_B$ (for $L>\xi$), a particle passes through regions already visited (``ergodic limit'', regime 3). Finally, for times much larger than the Heisenberg time $\tauh\equiv2\pi\overline{\nu} \text{min}(L,\xi)^2$ the particle has resolved the discreteness of energy levels and the system no longer evolves (``quantum limit'', regime 4).
} 
\label{phase_diagram}
\end{figure}

At lowest order in $(k_0\ell_B)^{-1}\ll1$, only pairs of trajectories following exactly the same multiple scattering sequence, i.e. with no net phase difference, survive the disorder average. The corresponding contribution to the kernel $\Phi_{\bk_0\bk E}(\omega)$, the so-called series of ladder diagrams, is shown in Fig. \ref{diagrams}(a). It describes a diffusion mechanism and leads to a fast isotropization process of atomic momenta at very short times $t\sim \taub$ where $\taub$ is the Boltzmann transport time. This is the regime 1 in Fig. \ref{phase_diagram} and it has been analyzed in \cite{Plisson13} by means of a kinetic approach. In the following we will not consider this regime, focusing only on times $t\gg\taub$ where the isotropization process is completed and diffusion is fully established (regime 2 in Fig. \ref{phase_diagram}). 
For $t\gg\taub$, the $\omega-$dependence of the diagram in Fig. \ref{diagrams}(a) is purely controlled by its central (ladder) part, given by the diffusion propagator $P_E(\bq,\omega)=1/(-i\omega+D_B\bq^2)$ (with $D_B$ the Boltzmann diffusion constant) at zero momentum \cite{Cherroret12}:
\begin{eqnarray}
\Phi_{\bk_0\bk E}(\omega)&=& 
2
\av{\langle\bk_0|\text{Im}\hat{G}^R(E)|\bk_0\rangle}\times
2\av{\langle\bk|\text{Im}\hat{G}^R(E)|\bk\rangle}
\nonumber\\
&&\times P_E(0,\omega)/[2\pi\overline{\nu}(E)],
\end{eqnarray}
where the two average Green's operators come from the `legs' of the diagram and where $\overline{\nu}(E)$ is the average density of states per unit volume at energy $E$.
Inserting this expression into Eq. (\ref{rhokprimet}) and performing the Fourier integral over $\omega$, we obtain the time-independent, isotropic and ring-shaped contribution to the disorder-averaged momentum distribution, well visible in Fig. \ref{twinpeaks}:
 \begin{equation}\label{Background}
\av{n}_\text{D}(\textbf{k}) =\av{n}_\text{D}(k)=\intdone{E}\dfrac{A(\bk,E)A(\bk_0,E)}{2\pi\overline{\nu}(E)},
\end{equation}
where we have introduced the spectral function $A(\bk,E)=2\pi\av{\langle\bk|\delta(E-\hat{H})|\bk\rangle}=-2\av{\langle\bk|\text{Im}\hat{G}^R(E)|\bk\rangle}$ \cite{Shapiro12}. Note that $\av{n}_\text{D}(\textbf{k})$ is indeed isotropic because the spectral function only depends on $k\equiv |\bk|$ \cite{Shapiro12}. 
The physical interpretation of Eq.~(\ref{Background}) is rather clear: the spectral function $A(\bk_0,E)$ describes
the probability density that the initial state with momentum $\bk_0$ has an energy $E$, while $A(\bk,E)/[2\pi\overline{\nu}(E)]$ in turn
describes the probability density that a state with energy $E$ has a momentum $\bk.$ Note that one could imagine a slightly different
situation where the initial state is not a plane wave, for exemple a wave packet with finite size,
see Sec.~\ref{feasibility}, or a more complicated state obtained after the disordered potential is progressively switched on.
The analysis developed in this paper can be used to describe such a variant, simply by replacing $A(\bk_0,E)$ by the
energy distribution of the initial state.

The series of crossed diagrams (b) in Fig. \ref{diagrams} gives a correction to Eq. (\ref{Background}) that describes the CBS peak growing around the backscattering direction $\bk=-\bk_0$. Its calculation follows the same lines as that of diagram (a) \cite{Cherroret12}. Exactly at backscattering $\bk=-\bk_0$, the CBS contribution reaches rapidly a stationary value given by
\begin{equation}\label{CBS_diag}
\av{n}_\text{CBS}(-\bk_0)=\intdone{E}\dfrac{A^2(\bk_0,E)}{2\pi\overline{\nu}(E)}.
\end{equation}
According to Eqs. (\ref{Background}) and (\ref{CBS_diag}), at backscattering the momentum distribution is exactly twice the value of the isotropic background, $\overline{n}(-\bk_0)=\overline{n}_\text{D}(k_0)+\overline{n}_\text{CBS}(-\bk_0)=2\overline{n}_\text{D}(k_0)$, which is an emblematic signature of the CBS effect resulting from a plane-wave source \cite{Aegerter09}. Finally, it was shown in \cite{Karpiuk12, Micklitz14} that at short enough times, the leading contribution to the CFS peak is given by the diagrams (c) and (d) in Fig. \ref{diagrams}, which combine two successive crossed and ladder sequences. These diagrams are peaked in the forward direction, unlike other diagrams of the same order of magnitude in perturbation theory, which provide flat contributions to the momentum distribution \cite{Karpiuk12}. Diagram (d) is obtained from  diagram (c) by time-reversing one of the complex amplitudes. Since the system we consider has the time-reversal symmetry, both diagrams equally contribute, giving
\begin{eqnarray}\label{CFS_diag}
\av{n}_\text{CFS}(\bk_0)&\simeq&2\intdone{E}\intdone{\omega} e^{-i\omega t}\dfrac{A^2(\bk_0,E)}{[2\pi\overline{\nu}(E)]^2\tau_s}\times\nonumber\\
&&\intdtwo{\bq}A(\bq,E)P_E(\bq+\bk_0,\omega)^2.
\end{eqnarray}
In two dimensions, Eq. (\ref{CFS_diag}) yields $\av{n}_\text{CFS}(\bk_0,t)/\av{n}_\text{D}(k_0)\sim 1/(k_0\ell_B)$, which is a constant, small contribution for weak disorder \cite{Karpiuk12} (note that this result is different from the case of a one-dimensional or quasi one-dimensional geometry, for which the corresponding Eq. (\ref{CFS_diag}) gives $\av{n}_\text{CFS}(\bk_0,t)\sim\sqrt{t}$ \cite{Micklitz14}). This means that the CFS peak is hardly visible in the diffusive regime. As shown in \cite{Karpiuk12}, Eq. (\ref{CFS_diag}) is only valid at short enough times, when higher-order corrections to the CFS peak remain small. The question of the description of CFS at longer times, where such corrections cannot be neglected anymore, is the object of the next section.\\

\subsection{Theoretical description of the momentum distribution in the ergodic and quantum regimes}

For a 2D disordered system of size $L$ and characterized by a localization length $\xi$, Eqs. (\ref{Background}), (\ref{CBS_diag}) and (\ref{CFS_diag}) strictly speaking hold only in the diffusive regime where $\taub\ll t\ll \text{min}(L,\xi)^2/D_B$ (regime $2$ in Fig. \ref{phase_diagram}). When $L<\xi$, $L^2/D_B\equiv\taud$ is the so-called Thouless time, i.e. the typical time needed by an atom to reach the boundary of the system \cite{Thouless74}. In the opposite limit $\xi< L$, $\xi^2/D_B\equiv \tau_\text{loc}$ can be interpreted as the localization time, i.e. the time scale at which atoms become sensitive to Anderson localization. In both cases, as soon as $t>\taud$ or $\tau_\text{loc}$, atoms start to feel that they are confined in a finite region of space: this is the ergodic regime, see Fig. \ref{phase_diagram}. In the ergodic regime the system still evolves, until it eventually resolves  its spectrum (``quantum limit'', regime 4 in Fig. \ref{phase_diagram}). This happens at a time scale known as the Heisenberg time, $\tauh\equiv2\pi\overline{\nu}\text{min}(L,\xi)^2$, which is the inverse of the mean level spacing in a volume of size $L^2$ (when $L<\xi$) or $\xi^2$ (when $L>\xi$). Note that in two dimensions the ratios $\tau_H/\tau_D,\ \tau_H/\tau_\text{loc}\sim k_0\ell_B$ are very large in the weak-disorder limit assumed here, such that the ergodic regime is typically very broad.

In the ergodic regime, atoms pass again and again through spatial regions they have already explored. This phenomenon produces a a highly non-perturbative accumulation of interference, and many corrections to the CFS diagrams in Fig. \ref{diagrams}(c) and \ref{diagrams}(d) become relevant and come into play. At first sight, one might think of summing up all these corrections by directly ``chaining'' an arbitrary number of series of crossed and ladder diagrams. This approach seems however hopeless because of the rapid proliferation of the number of such corrections 
at longer and longer times (the Hikami boxes connecting the series can be dressed in many possible ways \cite{Karpiuk12}). Nevertheless, a description of the transport dynamics in the ergodic regime can be obtained from the supersymmetric nonlinear $\sigma$-model developed by Efetov \cite{Efetov99}. Within this approach, the intensity propagator has the general form
\begin{widetext}
\begin{equation} \label{Susy}
\Phi_{\bk_0\bk E}(\omega)=\dfrac{A(\bk_0,E)A(\bk,E)}{4L^2}\int DQ\left[Q_{15}(0)Q_{51}(0)+
Q_{11}(\bk-\bk_0)Q_{55}(\bk_0-\bk)+
Q_{35}(\bk+\bk_0)Q_{53}(-\bk-\bk_0)\right]e^{-F[Q]}.
\end{equation}
\end{widetext}
In Eq. (\ref{Susy}), $\int (...)DQ$ is a functional integral over a $8\times 8$ supermatrix $Q$ that fulfills the constraint $Q^2=1$, and $\Lambda=\text{diag}(\mathbb{1}_4,-\mathbb{1}_4)$. The elements of $Q$ are complex and Grassmann fields (the Hamiltonian here belongs to the orthogonal symmetry class). The action of the $\sigma$-model is $F[Q]=[\pi\overline{\nu}(E)/4]\text{Str}\int d^2\textbf{r}[-D(\boldsymbol\nabla Q)^2-2i\omega\Lambda Q]$, where Str denotes the supertrace. The elements of $Q$ are arranged in two retarded and advanced sectors describing the product of the two Green's functions involved in $\Phi_{\bk_0\bk E}$. These sectors are split in two sectors containing variables and their complex conjugate (pertaining to the time-reversal symmetry), themselves being split in two bosonic and fermionic sectors required to perform the disorder average \cite{Efetov99, Mirlin00}. 

At this stage, Eq. (\ref{Susy}) is general, with the only restriction that time should be larger than $\tau_B$ and disorder should be weak. After integration over $Q$ the three terms in the right-hand-side give rise to functions of $\bk$ which are respectively (from left to right) constant, peaked around $\bk=\bk_0$, and peaked around $\bk=-\bk_0$, and can thus \emph{a posteriori} be identified as the isotropic background of the distribution, the CFS and the CBS peaks. Due to the complicated manifold spanned by the matrix $Q$ however \cite{Efetov99, Mirlin00}, these functions can only be calculated in a few specific cases. In the next section, we focus on the limit $L\ll\xi$ where an exact result can be obtained for the CFS contrast in the ergodic regime $t>\tau_D$, and even in the long-time, quantum regime $t\gg\tau_H$, where the momentum distribution is expected to reach its final, stationary form.

Finally, it is worthwhile to note that whatever the ratio $L/\xi$ but at times $\taub\ll t\ll \text{min}(L,\xi)^2/D_B$ (diffusive regime), the field integrals over $Q$ can also be performed, using a perturbation theory around the high-frequency saddle point $Q=\Lambda$ \cite{Micklitz14}. The calculation of the three terms in Eq. (\ref{Susy}) in this limit then reproduces Eq. (\ref{Background}) for the background, Eq. (\ref{CBS_diag}) for the CBS peak at $\bk=-\bk_0$, and Eq. (\ref{CFS_diag}) for the CFS peak at $\bk=\bk_0$. This in particular confirms the conjecture that at short times the series of diagrams in Figs. \ref{diagrams}(c) and \ref{diagrams}(d) are indeed those responsible for the CFS effect \cite{Karpiuk12}.


\section{CFS in a limited volume ($L\ll\xi$)}
\label{sectionL}

\subsection{CFS in the ergodic and quantum regimes}

In this section, we assume $L\ll \xi$ and focus on times $t\gg \taud$ (regions $2$ and $3$ in Fig. \ref{phase_diagram}). Since $L\ll\xi$, the confinement effect leading to the CFS peak stems from the finite volume of the system. Its dynamics  can be accessed from Eq. (\ref{Susy}) by replacing the functional integral by a definite one and using the parametrization of the matrix $Q$ proposed by Efetov \cite{Efetov99} (``zero-dimensional'' approximation). With this strategy, the CFS contribution to Eq. (\ref{Susy}) yields, at $\bk=+\bk_0$,
\begin{equation}
\label{CFS_longt}
\av{n}_\text{CFS}(\bk_0,t)=\intdone{E}\dfrac{A^2(\bk_0,E)}{2\pi\overline{\nu}(E)}2\pi\overline{\nu}(E)L^2K_E(t),
\end{equation}
where 
\begin{eqnarray}
\label{KE_GOE}
&&2\pi\overline{\nu}(E)L^2K_E(t)=\\ 
&&
\left \{\begin{array}{l}
    (t/\tauh)[2-\ln(1+2t/\tauh)],\ \taud\ll t\leq\tauh \\
    2-(t/\tau_H)\ln[(1+2t/\tauh)/(2t/\tauh-1)],\ t\geq\tauh, \nonumber\\
\end{array}
\right.
\end{eqnarray}
with $\tauh\equiv2\pi\overline{\nu}(E)L^2$ the Heisenberg time associated with the system of size $L$. This result shows that starting from $t=\taud$, the CFS peak slowly increases until a few Heisenberg times as atoms keep exploring the volume of the system. For $t\gg \tauh$, $2\pi\overline{\nu}(E)K_E(t)\simeq 1$ and the CFS peak has reached its maximum. In this limit, the momentum distribution no longer evolves in time because atoms have resolved the discreteness of energy levels. It is interesting to note that the function $K_E(t)$, as given by Eq. (\ref{KE_GOE}), is nothing but the so-called form factor -- the Fourier transform of the correlation of density-of-states fluctuations -- of the Gaussian Orthogonal Ensemble of random matrices \cite{Mirlin00} (see also Sec. \ref{sectionxi}). This shows in particular that the CFS peak is intrinsically connected with the spectral properties of the disordered system \cite{Loon14}. In real space, the form factor also governs the 
dynamics of the ``mesoscopic echo effect'', i.e. the enhancement of the probability for a spatially narrow wave packet to return to the origin in the presence of disorder \cite{Prigodin94}. 

Finally, we mention for completeness that the CBS peak at $\bk=-\bk_0$ as well as the isotropic component of the distribution [the third and first terms in Eq. (\ref{Susy}), respectively] can also be derived from Eq. (\ref{Susy}) for $t>\taud$, using the zero-dimensional approximation. This calculation eventually leads to the same expressions given in Eqs. (\ref{Background}) and (\ref{CBS_diag}), signaling that these formulas in fact hold very generally, not only in the diffusive regime but also in the long-time limit $t>\taud$.


\subsection{Numerical simulations}
\label{numerics_WD}

\begin{figure}[h]
\includegraphics[width=1\linewidth]{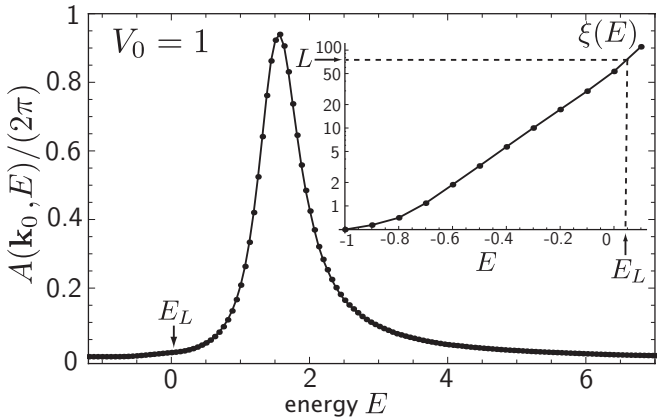}
\caption{
Spectral function $A(\bk_0,E)$ as a function of energy, obtained from numerical simulations of plane-wave propagation in a 2D speckle potential, for $V_0=1$ and $k_0=2$ (lengths, momenta and energies are respectively given in units of $\zeta$, $\zeta^{-1}$ and $1/(m\zeta^2)$, where $\zeta$ is the correlation length of the random potential). 
The inset shows the localization length given by the transfer-matrix approach as a function of energy, for the same value of $V_0$. 
Data are averaged over $16$ disorder realizations. The energy $E_L$ corresponding to a localization length $\xi(E_L)=L$ for $L=25\pi$ is indicated: energies below $E_L$ are typically localized [$\xi(E)<L$], while energies above $E_L$ are typically diffusive [$\xi(E)>L$]. From the main plot, it is seen that below $E=E_L$ the spectral function is almost zero, which means that for these parameters essentially all atoms are diffusive.} 
\label{SpectralF}
\end{figure}

For $L\ll\xi$, we now have a complete physical picture of the dynamics, with  Eqs. (\ref{Background}), (\ref{CBS_diag}) and (\ref{CFS_longt}) describing respectively the isotropic background, the CBS and the CFS peaks. In order to test the validity of these formulas, we perform extensive numerical simulations of the time-resolved momentum distribution of a matter wave in a 2D speckle potential, using the approach outlined in Sec. \ref{numerics}. To achieve the condition $L\ll\xi$, we set $k_0=2$ and consider a relatively weak value of the disorder amplitude, $V_0=1$. For these parameters, we compute numerically the spectral function $A(\bk_0,E)=-2\text{Im}\overline{\langle\bk_0|\hat{G}(E)|\bk_0\rangle}$, where $\langle\bk_0|\hat{G}(E)|\bk_0\rangle=-i\int_0^\infty dt\langle\bk_0|\text{exp}[i(E-\hat{H})t] |\bk_0\rangle$ is obtained by propagation of the plane-wave state $|\bk_0\rangle$. The spectral function is shown in the main panel of Fig. \ref{SpectralF}, giving an estimation of the energy distribution of the matter wave. Its shape is reminiscent of the Lorentzian expected in the limit of weak disorder \cite{AM}. The inset of Fig. \ref{SpectralF} also shows the energy dependence of the localization length, which we compute numerically using the transfer-matrix technique in two dimensions \cite{footnote1, McKinnon83, Slevin14, Delande14}. At a given $L$ corresponds a certain energy $E_L$, below which atoms are typically localized [$\xi(E<E_L)<L$] and above which atoms are typically diffusive [$\xi(E>E_L)>L$]. For the largest value of $L$ considered in this section ($L=25\pi$, see below) we find $E_L\simeq0.05$. This value falls in the left tail of the spectral function, where the latter is almost zero. 
This means that for $V_0=1$ and $k_0=2$ essentially all atoms fulfill the condition $L<\xi$. 

\begin{figure}
\includegraphics[width=0.9\linewidth]{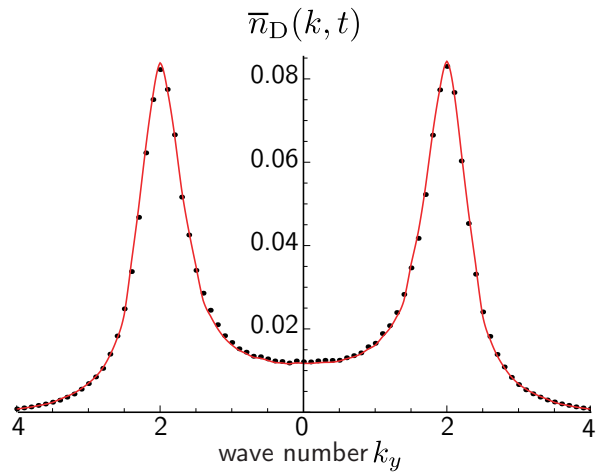}
\caption{(Color online) 
Cut along $k_x=0$ of the momentum distribution, describing the radial shape of its isotropic, ring-shaped part. Points were obtained from numerical simulations of plane-wave propagation in a 2D speckle potential in the limit $L\ll\xi$ ($V_0=1$ and $k_0=2$), after averaging over $6000$ disorder realizations. $L$ is set to $25\pi$. The red curve is the theoretical prediction (\ref{Background}), in which spectral functions and density of states are computed numerically. There is no adjustable parameter.} 
\label{background}
\end{figure}

We show in Fig. \ref{background} a radial cut along $k_x=0$ of the momentum distribution obtained numerically after a propagation time $t=7200$, i.e. well beyond the Boltzmann transport mean free time which is $\taub\simeq 7$ at $E=E_0\equiv k_0^2/(2m)$. This plot is expected to describe the radial shape of the isotropic part of the distribution, given by Eq. (\ref{Background}). In Fig. \ref{background} we also show this theoretical prediction, in which we used the numerically computed spectral functions $A(\bk_0,E)$, $A(\bk,E)$, and density of states $\overline{\nu}(E)\equiv\int d\bk/(2\pi)^3A(\bk,E)$. We see that the theory perfectly matches the numerical results without any adjustable parameter.

\begin{figure}
\centering\includegraphics[width=1\linewidth]{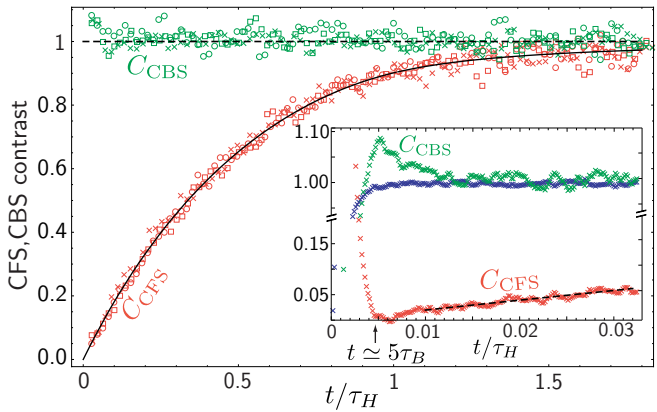}
\caption{(Color online) 
Main panel: contrast of the CBS (upper green symbols) and CFS (lower red symbols) peaks as a function of $t/\tauh$ [here $\tau_H=\tau_H(E_0)=2\pi\overline{\nu}(E_0) L^2$], obtained from numerical simulations of plane-wave propagation in the limit $L\ll\xi$ ($V_0=1$ and $k_0=2$). Circles were obtained for $L=15\pi$, squares for $L=20\pi$ and crosses for $L=25\pi$. Data are averaged over $6000$ disorder realizations and over a time window $\Delta t=250$. The dashed and solid curves are the theoretical predictions (\ref{CCBS}) and (\ref{CCFS}), respectively. Inset: contrasts at short times $t\ll\tau_H$. For comparison, the scaled background $\overline{n}_D(k_0,t)/\overline{n}_D(k_0,\infty)$ is also shown (blue crosses). Both the background and the CBS contrast become time independent after a few $\taub$, while the CFS contrast increases slowly, linearly in time.
} 
\label{CBSCFS_weak}
\end{figure}

Let us now focus on the contrast of the CBS and CFS peaks with respect to the isotropic background, defined respectively as $C_\text{CBS}\equiv \overline{n}_\text{CBS}(-\bk_0,t)/\overline{n}_\text{D}(k_0,t)$ and $C_\text{CFS}\equiv\overline{n}_\text{CFS}(\bk_0,t)/\overline{n}_\text{D}(k_0,t)$. According to Eqs. (\ref{Background}) and (\ref{CBS_diag}), we have evidently
\begin{equation}
\label{CCBS}
C_\text{CBS}=1,\ t\gg\taub.
\end{equation}
Furthermore, by comparing Eqs. (\ref{Background}) and (\ref{CFS_longt}) and assuming that the energy dependence of $\tauh$ is smooth as compared to that of the spectral function (which is a very good approximation for the relatively low value of $V_0$ considered in this section), we have
\begin{equation}
\label{CCFS}
C_\text{CFS}\simeq2\pi\overline{\nu}(E_0)L^2K_{E_0}(t),\ t\gg\taud,
\end{equation}
where the expression of $K_E(t)$ is given by Eq. (\ref{KE_GOE}). Eqs. (\ref{CCBS}) and (\ref{CCFS}) are shown in the main panel of Fig. \ref{CBSCFS_weak} as a function of time (dashed and solid curves, respectively), together with the CBS and CFS contrasts obtained from our numerical simulations of plane-wave propagation (green and red symbols, respectively). Circles were obtained for a system size $L=15\pi$, squares for $L=20\pi$ and crosses for $L=25\pi$. The agreement between analytical formulas and the numerics is excellent. Note in particular that all the numerical points fall on the same master curve when plotted as a function of $t/\tauh$, which confirms the universal scaling in $t/L^2$ of the function $K_{E_0}(t)$. The inset additionally shows the CBS and CFS contrasts together with the isotropic, background contribution to the momentum distribution at short times $t\ll\tauh$: both the  background and the CBS contrast become time independent after a few $\taub$, see Eqs. (\ref{Background}) and (\ref{CBS_diag}), while the CFS contrast increases slowly, linearly in time, in agreement with the small-time limit $2\pi\overline{\nu}(E_0)L^2K_{E_0}(t\ll \tauh)\simeq 2t/\tauh$.

\section{CFS in a localized system ($\xi\ll L$)}
\label{sectionxi}

\subsection{CFS peak at long times}

We now consider the case $\xi\ll L$, where the confinement is due to Anderson localization. This was the scenario originally studied in \cite{Karpiuk12}, and also the one pertaining to Fig. \ref{twinpeaks}. With respect to Sec. \ref{sectionL}, the localization time $\tau_\text{loc}\equiv \xi^2/D_B$ now plays the role of the Thouless time $\tau_D\equiv L^2/D_B$, and the Heisenberg time $\tauh=2\pi\overline{\nu} \xi^2$ now refers to a volume of size $\xi$. The case $\xi\ll L$ is extremely interesting since now \emph{the emergence of the CFS peak is a hallmark of localization}. When $\xi\ll L$, no exact solution of the nonlinear $\sigma$-model (\ref{Susy}) is unfortunately available in two dimensions, and consequently there is no exact expression for the CFS contrast for $t>\tau_\text{loc}$. Nevertheless, the time evolution of CFS can be estimated in some limiting cases. First, for $\tau_\text{loc}< t\ll \tauh$ (i.e. at the onset of the ergodic regime, see Fig. \ref{phase_diagram}),  it was 
suggested by a qualitative argument of renormalization of the diffusion coefficient in the diagrams in Fig. \ref{diagrams} that the CFS peak should increase as $t/\tauh$ \cite{Karpiuk12}. The isotropic background and the maximum of the CBS peak were on the other hand predicted to remain unchanged at any time $t\gg\taub$, namely to be still given by Eqs. (\ref{Background}) and (\ref{CBS_diag}), respectively \cite{Karpiuk12}. As we now show, the CFS contrast for $\xi\ll L$ can also be evaluated in the long-time limit $t\gg\tauh$ (quantum regime, see Fig. \ref{phase_diagram}). For this purpose, we first recognize that the function $K_E(t)$ in Eq. (\ref{CFS_longt}) that we derived for $L\ll\xi$ is nothing but the so-called form factor:
\begin{equation}
\label{form_factor}
K_E(t)=\intdone{\omega}e^{-i \omega t}K_E(\omega),
\end{equation}
where $K_E(\omega)=\overline{\delta\nu(E+\omega/2)\delta\nu(E-\omega/2)}/\overline{\nu}(E)^2$ is the Fourier transform of the correlation function of density-of-states fluctuations, which can be rewritten as \cite{Mirlin00}
\begin{eqnarray}
\label{level_cor}
&&K_E(\omega) = -1+\nonumber\\
&&\dfrac{1}{\overline{\nu}(E)^2L^4}
\overline{\sum_{i,j}\delta\left(E+\dfrac{\omega}{2}-E_i\right)\delta\left(E-\dfrac{\omega}{2}-E_j\right)},
\end{eqnarray}
where the $E_i$ are the energy levels of the disordered system. In fact, the relation (\ref{CFS_longt}) between the CFS peak and the form factor defined by (\ref{level_cor}) turns out to hold very generally, for any ratio of $L$ and $\xi$. This can be explicitly shown by a modal decomposition of the wave function written in momentum space, a task that was accomplished in \cite{Loon14}. With the help of this relation, the problem of calculating the CFS contrast as a function of time boils down to the analysis of the frequency dependence of the correlation function $K_E(\omega)$. In the localization regime $\xi\ll L$ and in the limit of small frequencies, this dependence can be accessed within a simple model of ``correlated localization volumes'', originally introduced by Mott \cite{Mott70}. Let us briefly recall the main lines of this model. We here essentially follow the point of view of \cite{Sivan87, Atland95}: we conceptually divide our 2D disordered system in small patches of volume $\xi^2$, such that in 
each patch the mean level spacing is $\Delta\equiv(\overline{\nu}\xi^2)^{-1}=2\pi/\tauh$. Within one patch, two eigenstates experience the usual level repulsion of disordered systems \cite{Mehta} and are therefore far apart in the spectrum. Conversely, let us consider two close levels $E_i$ and $E_j$ such that $|E_i-E_j|\equiv|\omega|\ll\Delta$. The corresponding eigenstates then belong to two distant localization patches. We model the sub-system formed by these two levels by the coupling Hamiltonian
\begin{equation}
\label{Hmatrix}
H_c=\begin{pmatrix}
\epsilon_1 & \Delta e^{-|\textbf{r}_1-\textbf{r}_2|/\xi} \\
\Delta e^{-|\textbf{r}_1-\textbf{r}_2|/\xi}  & \epsilon_2
\end{pmatrix}.
\end{equation}
Here $\epsilon_1$ and $\epsilon_2$ are the energy levels in the absence of coupling, and $\Delta e^{-|\textbf{r}_1-\textbf{r}_2|/\xi}$ is the overlap integral between the two uncoupled states, whose wave functions are exponentially localized around $\textbf{r}_1$ and $\textbf{r}_2$, respectively. Due to the coupling, the levels become $\bar\epsilon\pm\sqrt{\delta\epsilon^2/4+\Delta^2e^{-2r/\xi}}$, where $\bar\epsilon\equiv(\epsilon_1+\epsilon_2)/2$, $\delta\epsilon\equiv\epsilon_1-\epsilon_2$ and $r\equiv|\textbf{r}|\equiv|\textbf{r}_1-\textbf{r}_2|$. As in \cite{Sivan87}, we assume that $\bar\epsilon$, $\delta\epsilon$ and $\textbf{r}$ are independent, uniformly distributed random variables, respectively over an interval of size $\Delta$ (for $\bar\epsilon$ and $\delta\epsilon$) and over the volume $L^2$ (for $\textbf{r}$). Performing the integral over $\bar\epsilon$ allows us to get rid of one of the delta functions in Eq. (\ref{level_cor}), which yields
\begin{equation}
\label{K_eq}
K_E(\omega)\sim\int_{-\Delta/2}^{\Delta/2}d\delta\epsilon
\int \dfrac{d^2\textbf{r}}{L^2}\ 
\delta\left(\omega-\sqrt{\delta\epsilon^2+4\Delta e^{-2r/\xi}}\right).
\end{equation}
In Eq. (\ref{K_eq}), the integral over the difference of localization centers ranges over the full volume of the system, $L^2$. The bounds in the integral over $\delta\epsilon$ account for the fact that the absolute difference between $\epsilon_1$ and $\epsilon_2$ should not be greater that $\Delta$ because we only consider the coupling between states belonging to different localization patches. The two integrals are readily performed, using the inequality $|\omega|\ll\Delta$ and taking the limit $\xi\ll L$. We obtain
\begin{equation}
\label{Komega_final}
K_E(\omega)\sim\left(\dfrac{\xi}{L}\right)^2\ln^2\left(\dfrac{|\omega|}{2\Delta}\right),\ |\omega|\ll \Delta.
\end{equation}
Note that in deriving Eq. (\ref{Komega_final}), we implicitly assumed $\omega\ne0$. In order to describe long times, we must also include the contribution $\omega=0$, which comes from the diagonal terms $i=j$ in the sum in Eq. (\ref{level_cor}). These terms yield the contribution $\delta(\omega)/[\overline{\nu}(E) L^2]$, which describes the self correlation of one energy level. Adding it to Eq. (\ref{Komega_final}) and performing the Fourier transform with respect to $\omega$, we obtain the final form of $K_E(t)$ and of the CFS contribution to the momentum distribution at $\bk=+\bk_0$, at long times $t\gg\tauh$:
\begin{equation}
\label{CFS_longt_loc}
\av{n}_\text{CFS}(\bk_0,t)\simeq\intdone{E}\dfrac{A^2(\bk_0,E)}{2\pi\overline{\nu}(E)}\left[1-\alpha\dfrac{\ln(\beta t/\tauh)}{t/\tauh}\right].
\end{equation}
In writing Eq. (\ref{CFS_longt_loc}), we have introduced two phenomenological parameters, $\alpha$ and $\beta$, whose precise determination is not accessible from the present approach. $\alpha$ accounts for the fact that in a real system, the distributions of  $\bar\epsilon$, $\delta\epsilon$ and $r$ may not be exactly uniform, while $\beta$ accounts for the fact that the strength of the coupling terms in the Hamiltonian (\ref{Hmatrix}) may slightly differ from $\Delta$. The appearance of a logarithm in Eq. (\ref{CFS_longt_loc}) is however typical of 2D disordered systems \cite{Atland95}. This has to be contrasted with one-dimensional or quasi one-dimensional geometries for which $K_E(\omega)\propto\ln[|\omega|/(2\Delta)]$, leading eventually to a purely algebraic decay $\propto(t/\tauh)^{-1}$ of the second term in the right-hand-side of Eq. (\ref{CFS_longt_loc})  \cite{Micklitz14, Loon14}.

\subsection{Numerical simulations for $\xi\ll L$}
\label{numerics_loc}

\begin{figure}
\includegraphics[width=1\linewidth]{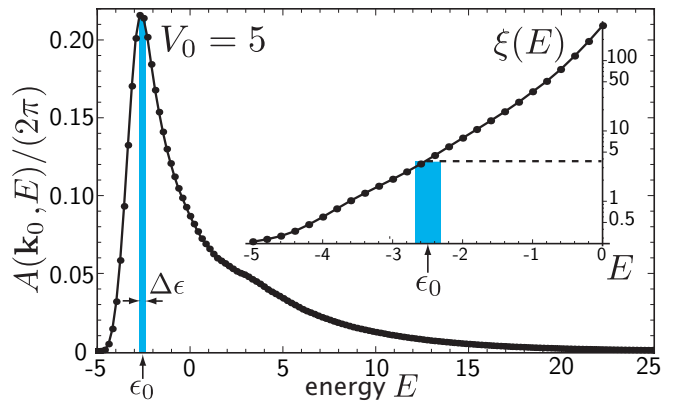}
\caption{(Color online) 
Spectral function $A(\bk_0,E)$ as a function of energy, for $V_0=5$ and $k_0=1.5$ (lengths, momenta and energies are respectively given in units of $\zeta$, $\zeta^{-1}$ and $1/(m\zeta^2)$, where $\zeta$ is the correlation length of the random potential). For the time evolution, only energies in the narrow band $[\epsilon_0-\Delta\epsilon/2,\epsilon_0+\Delta\epsilon/2]$ are selected, where $\epsilon_0=-2.5$ and $\Delta\epsilon=0.4$. 
The inset shows the localization length given by the transfer-matrix approach as a function of energy, for the same value of $V_0$. 
Data are averaged over $16$ disorder realizations.} 
\label{SpectralF5}
\end{figure}

Let us now confront the predictions (\ref{Background}), (\ref{CBS_diag}) and (\ref{CFS_longt_loc}) with numerical simulations. To describe the localization regime, it is necessary to make the localization length smaller than the system size, and it is thus mandatory to use a stronger value of the disorder amplitude $V_0$. The main panel of Fig. \ref{SpectralF5} displays the numerically computed spectral function $A(\bk_0,E)$ for $V_0=5$ and $k_0=1.5$.
In sharp contrast with Sec. \ref{numerics_WD} where disorder was relatively weak, the spectral function has now a maximum at negative energy and a long tail toward high energies. This raises two new problems. First, for a given system size $L$ it is hard to fulfill the inequality $\xi(E)\ll L$ for all energies. For instance, when $L=100\pi$,  the energy $E_L$ such that $\xi(E_L)=L$ is slightly above zero, thus distinctly above the maximum of the spectral function, see the main panel of Fig. \ref{SpectralF5}. Therefore, many particles have $E>E_L$ and thus have a localization length $\xi(E)> L$. Second, unlike in Sec. \ref{numerics_WD} the Heisenberg time (which is proportional to the square of the localization length) now varies much faster with $E$ than the spectral function. As a consequence, it is not even possible to identify a single Heisenberg time for localized atoms.
In order to nevertheless consider a ``clean'' situation where all atoms are localized with approximately the same Heisenberg time,
we introduce a filtering in energy in the time-propagation algorithm: at $t=0$ we apply the operator $\exp[-(\hat{H}-\epsilon_0)^2/(2\Delta\epsilon^2)]$ to the initial state $|\bk_0\rangle$, and only then propagate it with the evolution operator $\exp(-i\hat{H}t)$. With this procedure the energies $E$ involved in transport roughly lie in the interval $[\epsilon_0-\Delta\epsilon/2,\epsilon_0+\Delta\epsilon/2]$, which is chosen so that $\xi(E)\ll L$ for all $E$ within that interval. In the rest of this section, we choose $\epsilon_0=-2.5$ and $\Delta\epsilon=0.4$ for $V_0=5$ and $k_0=1.5$, see Fig. \ref{SpectralF5}. With these parameters, the localization length given by the transfer-matrix approach at $E=\epsilon_0$ is $\xi\simeq 3.9$. From a theoretical point of view, the filtering in energy amounts to performing the replacement $A(\bk, E)\rightarrow A(\bk, E)\exp\left[-(E-\epsilon_0)^2/(2\Delta\epsilon^2)\right]$ (for both $\bk$ and $\bk_0$) in all formulas.
\begin{figure}
\includegraphics[width=0.9\linewidth]{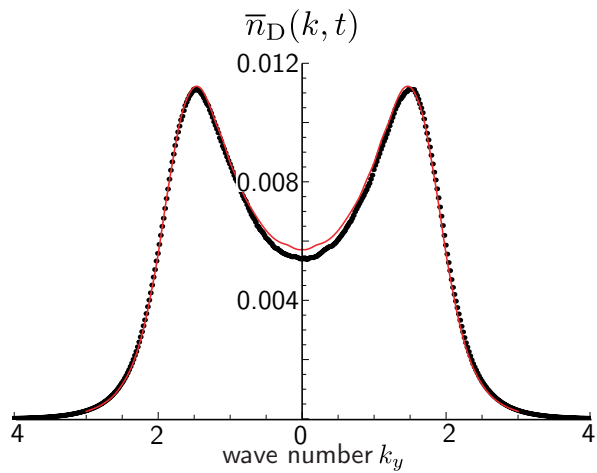}
\caption{(Color online) 
Cut along $k_x=0$ of the momentum distribution, describing the radial shape of its isotropic, ring-shaped part. Points were obtained from numerical simulations of plane-wave propagation in a 2D speckle potential in the limit $\xi\ll L$ ($V_0=5$, $k_0=1.5$, $\epsilon_0=-2.5$, $\Delta\epsilon=0.4$), after averaging over $240$ disorder realizations. $L$ is set to $100\pi$. The red curve is the prediction (\ref{Background}), in which spectral functions and density of states are computed numerically. There is no adjustable parameter.} 
\label{background_v5}
\end{figure}

We first show in Fig. \ref{background_v5} a radial cut along $k_x=0$ of the numerically computed momentum distribution (black points). These points are expected to describe the radial shape of the isotropic background. The chosen time is $t=1500$, i.e. well beyond the transport mean free time which is $\taub\simeq 2.6$. In the same plot we show the theoretical prediction (\ref{Background}) fed with the numerically computed spectral functions  $A(\bk_0,E)$, $A(\bk,E)$, and density of states $\av{\nu}(E)$.
The agreement is very good and may come as a surprise if we remember that Eq. (\ref{Background}) has been actually derived in the weak-disorder limit. It demonstrates the general validity of Eq. (\ref{Background}) for the isotropic background, even in the deep localization regime $\xi\ll L$ and for rather strong disorder, provided the \emph{exact} $A(\bk,E)$ and $\av{\nu}(E)$ are used in the computation. This suggests that at strong disorder, all interference corrections to the isotropic background boil down to a renormalization of the scattering mean free path while the global topology of the diagram in Fig. \ref{diagrams}(a) remains valid. Note in passing that the dip around $k_y=0$ is less pronounced in Fig. \ref{background_v5} than in Fig. \ref{background}. This is  a direct consequence of the long energy tail of the spectral function at stronger disorder.
\begin{figure}[h]
\centering\includegraphics[width=1\linewidth]{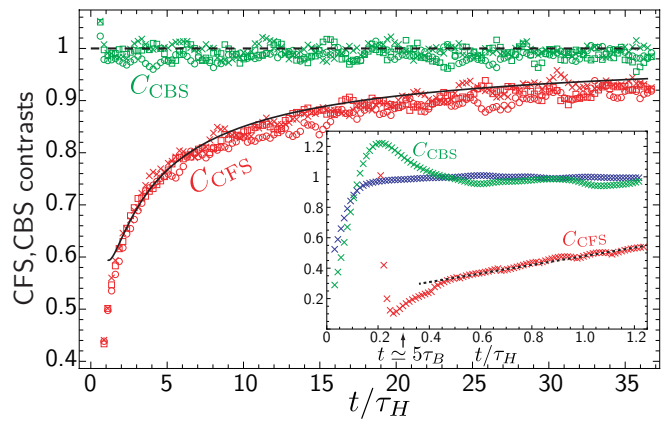}
\caption{(Color online) 
\label{CBSCFS_strong}
Main panel: contrast of the CBS (upper green symbols) and CFS (lower red symbols) peaks as a function of $t/\tauh$, obtained from numerical simulations of plane-wave propagation in the localization regime $\xi\ll L$, for the same parameters as in Fig. \ref{background_v5} (here $\tauh\simeq40$ \cite{footnote2}). Circles were obtained for $L=50\pi$, squares for $L=80\pi$ and crosses for $L=100\pi$. Data are averaged over $240-1600$ disorder realizations (depending on the value of $L$) and over a time window $\Delta t=40$. The dashed and solid curves are the theoretical predictions (\ref{Contrast_CBS_loc}) and (\ref{Contrast_CFS_loc}), respectively. Inset: contrasts at short times $t\ll\tauh$. For comparison, the scaled background $\overline{n}_D(k_0,t)/\overline{n}_D(k_0,\infty)$ is also shown (blue crosses). Both the background and the CBS contrast become time independent after a few $\taub$, while the CFS contrast increases slowly, linearly in time, as highlighted by the dotted line.
} 
\end{figure}

\begin{figure}
\includegraphics[width=0.95\linewidth]{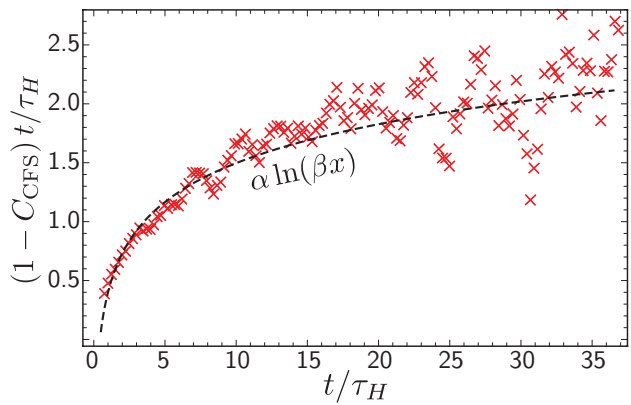}
\caption{(Color online) 
\label{CFS_long}
Red crosses: numerical values of $(1-C_\text{CFS})t/\tauh$ plotted as a function of $t/\tauh$, for the same parameters as in Fig. \ref{CBSCFS_strong}, with $L$ set to $100\pi$. The dashed curve is the function $\alpha\ln(\beta x)$, where $\alpha=0.5$ and $\beta=2.3$.
Data are averaged over $240$ disorder realizations and over a time window $\Delta t=40$.
}
\end{figure}

The CBS contrast for $\xi\ll L$ is still given by the ratio of Eqs. (\ref{Background}) and (\ref{CBS_diag}), and has thus the same expression as in Sec. \ref{sectionL}:
\begin{equation}
\label{Contrast_CBS_loc}
C_\text{CBS}=1,\ t\gg\taub.
\end{equation}
On the other hand, the CFS contrast at times $t\gg\tauh$ now follows from Eq.  (\ref{CFS_longt_loc}):
\begin{equation}
\label{Contrast_CFS_loc}
C_\text{CFS}=1-\alpha\dfrac{\ln(\beta t/\tauh)}{t/\tauh},\ t\gg\tauh.
\end{equation}
These two relations are shown in the main panel of Fig. \ref{CBSCFS_strong} (dashed and solid curves, respectively) together with the CBS and CFS contrasts obtained from our numerical simulations (green and red symbols, respectively), for $\tauh\simeq 40$ \cite{footnote2}. Circles were obtained for a system size $L=50\pi$, squares for $L=80\pi$ and crosses for $L=100\pi$. We find that Eq. (\ref{Contrast_CFS_loc}) well reproduces the numerical results for $\alpha=0.5\pm 0.1$ and $\beta=2.3\pm 0.1$ [for the fit we only consider times larger than $\tauh$, which is the limit of validity of Eq. (\ref{CFS_longt_loc})]. Note that as opposed to the case $L\ll\xi$ (see Fig. \ref{CBSCFS_weak}), the contrast of the CFS no longer depends on $L$, as expected in the localization regime.  The inset additionally shows the CBS and CFS contrasts together with the background contribution to the momentum distribution at short times $t\ll\tauh$. The observed behavior is qualitatively the same as in the case $L\ll\xi$: both the background and CBS contrast become 
time independent after a few $\taub$, while the CFS contrast increases slowly in time. This increase is compatible with the linear law estimated in \cite{Karpiuk12}, though the latter is seen in a rather small time interval. In order to have a better estimation of the validity of the theoretical expression for the CFS contrast at long times, Eq. (\ref{Contrast_CFS_loc}), we replot in Fig. \ref{CFS_long} the quantity $(1-C_\text{CFS})t/\tauh$ as a function of $t/\tauh$. In this representation, the numerical points increase logarithmically in time, in full agreement with Eq. (\ref{Contrast_CFS_loc}).

\begin{figure}[h]
\includegraphics[width=0.9\linewidth]{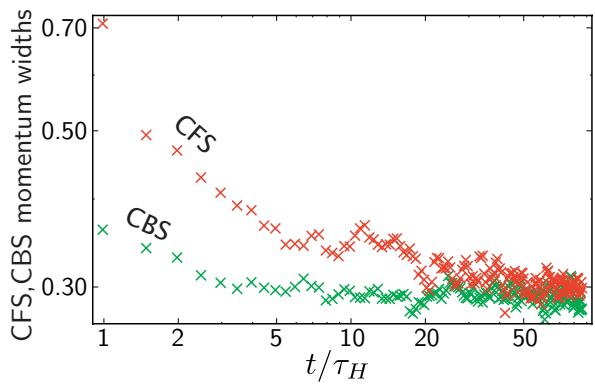}
\caption{(Color online) 
\label{CFS_width}
Width of the CBS and CFS peaks in momentum space as a function of $t/\tauh$, in the localization regime $\xi\ll L$, for the same parameters as in Fig. \ref{CBSCFS_strong}. $L$ is set to $100\pi$.
At long times, both widths converge to the same constant value, which is of the order of a few $\xi^{-1}$.  Data are averaged over $240$ disorder realizations and over a time window $\Delta t=40$.
}
\end{figure}

For the sake of completeness, we finally show in Fig. \ref{CFS_width} a plot of the width at half maximum in momentum space of both the numerical CBS and CFS peaks as a function of time. As for the contrast, we see that the evolutions of the two peaks are different. Over a time scale of the order of the mean free time, the CBS width quickly converges to a value of the order of a few $\xi^{-1}$. The CFS width, on the other hand, slowly decreases, until it reaches the same value as the CBS width after a few Heisenberg times, suggesting identical CBS and CFS profiles at very long times.

\section{Experimental scenario}
\label{feasibility}

In Sec. \ref{sectionxi}, we showed that when $\xi\ll L$ the CFS peak is a signature of Anderson localization, which confines atoms in a region of size $\xi$. However, we considered an ideal scenario where: (i) the dynamics is supported by a single energy $E=\epsilon_0$ and (ii) the initial state is a plane wave. In current state-of-the-art experiments on ultracold atoms however, these two conditions are not fulfilled. Indeed, all the energy components authorized by the spectral function shown in Fig. \ref{SpectralF5} contribute to transport, which means in particular that both diffusive and localized atoms are present. 
In addition, the initial state is never a plane wave but rather a wave packet of finite size $(\Delta k)^{-1}\ne\infty$ in configuration space. It thus remains important to clarify whether, within this non-ideal scenario, the CFS peak due to localized atoms is visible or not. This is the object of the present section.

\subsection{Effect of a broad energy distribution}

Let us first address the effect of a broad energy distribution on the CFS dynamics. For this purpose, as in Sec. \ref{numerics_loc} we numerically carry out the time evolution of a plane wave [$(\Delta k)^{-1}=\infty$] for $V_0=5$ and $k_0=1.5$, but this time \emph{without} applying the filtering in energy,  
\begin{figure}[h]
\includegraphics[width=0.95\linewidth]{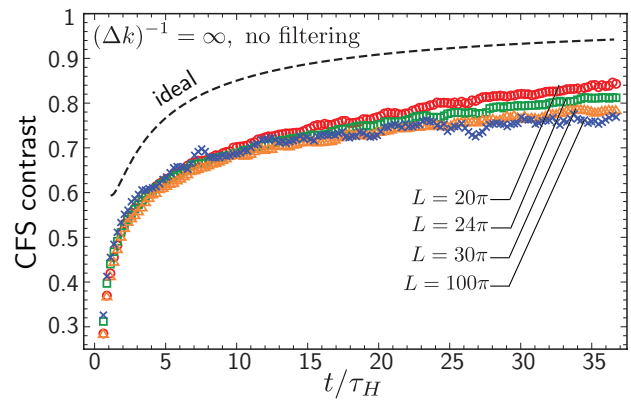}
\caption{(Color online) 
\label{CFS_L_effect}
Colored symbols: contrast of the CFS peak as a function of time, obtained from numerical simulations of plane-wave propagation [$(\Delta k)^{-1}=\infty$] for $V_0=5$ and $k_0=1.5$, without using any filtering in energy (lengths, momenta, energies and times are respectively given in units of $\zeta$, $\zeta^{-1}$, $1/(m\zeta^2)$ and $m\zeta^2$, where $\zeta$ is the correlation length of the random potential). The four curves correspond to different values of $L$. Data are averaged over $400$ disorder realizations and over a time window $\Delta t=40$. For comparison, the dashed curve shows the theoretical prediction (\ref{Contrast_CFS_loc}), corresponding to the ideal case of a single energy component $E=\epsilon_0=-2.5$, for which  $\tauh\simeq 40$. 
}
\end{figure}
such that now both localized and diffusive atoms coexist. The contrast of the CFS peak obtained in this way is shown in Fig. \ref{CFS_L_effect} as a function of time, for four values of the system size $L$. Several observations can be made. First, when $L$ is small, the CFS contrast decreases with $L$. This effect is due to diffusive atoms, which fulfill $L<\xi(E)$ and thus produce a CFS peak because of their confinement in the volume $L^2$ (mechanism discussed in Sec. \ref{sectionL}). Second, for the parameters used in Fig. \ref{CFS_L_effect}, the CFS contrast no longer visually changes with $L$ when $L\gtrsim 50\pi$. This means that in this limit -- which effectively corresponds to the experimental scenario $L=\infty$ -- the observed CFS peak is \emph{entirely due to localized atoms} (at smaller $L$ this is also the case at short enough times). Overall, the CFS contrast is however smaller than the ideal situation of Sec. \ref{sectionxi} where the dynamics was supported by a single, localized energy [Eq. (\ref{Contrast_CFS_loc}), dashed curve in Fig. \ref{CFS_L_effect}], because at a given time $t$ many localized atoms 
have an Heisenberg time $\tauh(E)>t$ and thus have not yet contributed to the CFS peak.

\subsection{Effect of the size of the wave packet}

\begin{figure}[h]
\includegraphics[width=0.95\linewidth]{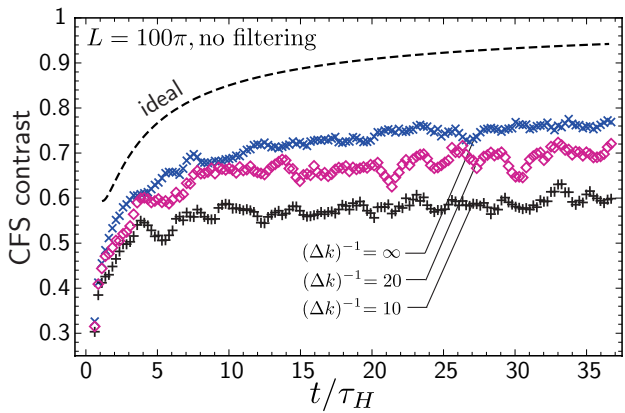}
\caption{(Color online) 
\label{CFS_Deltak_effect}
Colored symbols: contrast of the CFS peak as a function of time, obtained from numerical simulations starting from a Gaussian wave packet of finite width $(\Delta k)^{-1}$, for $V_0=5$, $k_0=1.5$ and $L=100\pi$, without using any filtering in energy. The three curves correspond to different values of $(\Delta k)^{-1}$, increasing from bottom to top. Data are averaged over $400$ disorder realizations and over a time window $\Delta t=40$. For comparison, the dashed curve shows the theoretical prediction (\ref{Contrast_CFS_loc}), corresponding to the ideal case of a single energy component $E=\epsilon_0=-2.5$, for which $\tauh\simeq 40$. 
}
\end{figure}

Having discussed the effect of a broad energy distribution, we now additionally consider the effect of the finite size of the initial wave packet. We show in Fig. \ref{CFS_Deltak_effect} the contrast of the CFS peak as a function of time for $L=100\pi$, obtained from numerical simulations starting from a Gaussian wave packet of width $(\Delta k)^{-1}\ne\infty$ rather than from a plane wave (as before $V_0=5$ and $k_0=1.5$ and no filtering in energy is applied). The finite value of $(\Delta k)^{-1}$ leads to a decay of the CFS contrast, well visible in the figure. This phenomenon can be traced back to Eq. (\ref{rhokprimet}): using a Gaussian wave packet amounts to taking $n_0(\bk')=(4\pi/\Delta k^2)\exp[-(\bk'-\bk_0)^2/(\Delta k^2)]$ instead of $(2\pi)^2\delta(\bk'-\bk_0)$, and thus to convolving the CFS peak obtained for a plane wave with a Gaussian function. From a physical point of view, the finite size of the wave packet cuts multiple scattering trajectories whose start and end points are separated by more than $(\Delta k)^{-1}$, as 
for the CBS effect \cite{Cherroret12}.

\section{Summary and concluding remarks}
\label{Conclusion}

This paper was devoted to a systematic study of the momentum distribution of a matter wave launched with finite velocity in a 2D random, speckle potential. In particular, we analyzed in detail the slowly evolving coherent forward scattering peak arising in the momentum distribution at long enough times. We showed that the emergence of this peak is conditional on the presence of a mechanism of confinement which allows to enhance the interference mechanism scattering particles in the forward direction. This confinement can arise in the situation where particles propagate diffusively in a bounded volume, or because of Anderson localization, which prevents transport beyond scales of the order of the localization length $\xi$. We studied both numerically and theoretically the two limits $L\ll\xi$ and $\xi\ll L$, for which we summarize the asymptotic expressions for the CFS contrast in Table \ref{table1}, for the ideal case where transport is supported by a single energy component.
\begin{table}[h]
\begin{tabular}{>{\centering}p{3.7cm}|>{\centering}p{2.2cm}|>{\centering}p{2.3cm}}
time &
$L\ll\xi$
$\tauh\equiv 2\pi\overline{\nu}L^2$
&
$\xi\ll L$ 
$\tauh\equiv 2\pi\overline{\nu}\xi^2$
\tabularnewline
\hline\hline
Diffusive regime
$\taub\ll t\ll\text{min}(L,\xi)^2/D_B$
&
$\propto \dfrac{1}{k_0\ell_B}$&
$\propto \dfrac{1}{k_0\ell_B}$
\tabularnewline
\hline
Ergodic regime
$\text{min}(L,\xi)^2/D_B\ll t\ll\tauh$
&
$2\dfrac{t}{\tauh}$&
$\propto\dfrac{t}{\tauh}$
\tabularnewline
\hline
Quantum regime\\
$t\gg\tauh$
&
$1-\dfrac{1}{12 (t/\tauh)^2}$&
$1-\alpha\dfrac{\ln(\beta t/\tauh)}{t/\tauh}$
\tabularnewline
\end{tabular}
\caption{\label{table1} Summary of asymptotic expressions for the CFS contrast $C_\text{CFS}$ in a 2D disordered system. $\alpha=0.5\pm 0.1$ and $\beta=2.3\pm 0.1$.}
\end{table}
From our results, it thus turns out that CFS of a matter wave could be experimentally observed by either artificially confining atoms in some finite regime of space, for instance with an optical potential with steep enough edges, or, more interestingly, by achieving Anderson localization. In current experimental setups, for instance that in \cite{Josse12}, atomic motion is not limited by any artificial boundary in the relevant directions of propagation, which corresponds to $L=\infty$. Consequently, any observation of the CFS effect using those setups \emph{would be a genuine signature of Anderson localization}. In an experiment, the visibility of CFS can be reduced because the matter wave supports many energy components and has initially a finite spatial width, but we showed that this reduction of visibility is rather small.

From a theoretical point of view, we saw that, except for short times, the dynamics of the CFS peak is not accessible from perturbation theory and requires the help of the nonlinear $\sigma$-model or of another non-perturbative approach. Still, in two dimensions there is presently no exact solution of the $\sigma$-model for $\xi\ll L$, such that no exact expression of the CFS peak is available at all times in this limit. This conclusion also applies to the three-dimensional case where the Anderson transition is present, which offers an interesting theoretical challenge for future works.

In the search for the CFS peak, ultracold atoms have many advantages, including the possibility for \emph{in situ} measurements of the velocity distribution. In principle however, the CFS peak could be also observed with other types of waves and in particular with classical waves propagating in disordered media. In this context, experiments often involve a ``scattering setup'' in which the wave is sent from outside the disordered system and transport properties are probed in transmission or reflection. While the reflection is well known to exhibit a prominent CBS peak \cite{Bayer93, Tourin97, Wolf85, Albada85}, the possibility of observing a CFS peak in the transmitted profile is more speculative. Indeed, in this setup both a high signal-to-noise ratio and a good angular resolution would be required to detect the CFS peak, the latter being very narrow and sitting on top of an exponentially small transmission signal. Closer to the situation described in the present work on the other hand, the scenario of 
transverse localization of light in paraxial geometries seems more promising \cite{Raedt89, Schwartz07}.

Coming back to the atomic context, it would finally be interesting to see how the CBS and CFS dynamics are perturbed by the presence of weak interactions between atoms, typically described by the nonlinear Gross-Pitaevskii equation for bosons. For a wave packet expanding in disorder, a weak nonlinearity is known to partially destroy Anderson localization and to restore a transport slower than diffusion \cite{Cherroret14, Kopidakis08, Pikovsky08, Basko11}. Nonlinearity-driven subdiffusion could manifest as well in momentum space and alter the CBS and CFS peaks, as what is known to happen to CBS in stationary setups \cite{Hartung08}.


\section*{Acknowledgements}

NC thanks Cord M\"uller for useful advice and comments, and the hospitality of the CQT, where part of this work was carried out. SG, CM and DD thank the Institut Fran\c{c}ais de Singapour (French Ministry of Foreign Affairs) and the National University of Singapore for supporting this work through the Merlion Programme. The authors acknowledge financial support from the French agency ANR (Project No. 11-B504-0003 LAKRIDI). This work was granted access to the HPC resources of TGCC under the allocation 2014-057083 made by GENCI (Grand Equipement National de Calcul Intensif) and to the HPC resources of The Institute for scientific Computing and Simulation financed by Region Ile de France and the project Equip@Meso (reference ANR-10-EQPX-29-01). The Centre for Quantum Technologies is a Research Centre of Excellence founded by the Ministry of Education and the National Research Foundation.


\begin{thebibliography}{99}


\bibitem{Anderson58} 
P.~W.~Anderson, Phys. Rev. \textbf{109}, 1492 (1958).

\bibitem{Bergmann84}
G. Bergmann, Phys. Rep. \textbf{107}, 1 (1984).

\bibitem{Hu08}
H. Hu, A. Strybulevych, J. H. Page, S. E. Skipetrov, and B. A. van Tiggelen.
Nature Physics \textbf{4}, 945 (2008).

\bibitem{Schwartz07}
T. Schwartz, G. Bartal, S. Fishman, and M. Segev,
Nature \textbf{446}, 52 (2007).

\bibitem{Maret12}
T. Sperling, W. B\"{u}hrer, C. M. Aegerter and G. Maret,
Nature Photonics \textbf{7}, 48 (2012).

\bibitem{Chabe08}
J. Chab\'e, G. Lemari\'e, B. Gr\'emaud, D. Delande, P. Szriftgiser, and J. C. Garreau,
Phys. Rev. Lett. \textbf{101}, 255702 (2008). 

\bibitem{Lemarie10}
G. Lemari\'e, H. Lignier, D. Delande, P. Szriftgiser, and J. C. Garreau,
Phys. Rev. Lett. \textbf{105}, 090601 (2010).

\bibitem{Billy08}
J. Billy, V. Josse, Z. Zuo, A. Bernard, B. Hambrecht,  P. Lugan, D. Cl\'ement, L. Sanchez-Palencia, P. Bouyer, and Aspect,
Nature \textbf{453}, 891 (2008).
  
\bibitem{Roati08}
G. Roati, C. D'\'Errico, L. Fallani, M. Fattori, C. Fort, M Zaccanti, G. Modugno, M. Modugno and M. Inguscio,
Nature \textbf{453}, 895 (2008).

\bibitem{Kondov11}
S. S. Kondov, W. R. McGehee, J. J. Zirbel, and B. DeMarco, Science \textbf{334}, 66 (2011).

\bibitem{Jendrzejewski12}
F. Jendrzejewski, A. Bernard, K. M\"uller, P. Cheinet, V. Josse, M. Piraud, L. Pezz\'e, L. Sanchez-Palencia, A. Aspect, and P. Bouyet,
Nature Phys. \textbf{8}, 398 (2012).

\bibitem{Semeghini14}
G. Semeghini, M. Landini, P. Castilho, S. Roy, G. Spagnolli, A. Trenkwalder, M. Fattori, M. Inguscio, and G. Modugno,
arXiv:1404.3528.

\bibitem{Cherroret12}
N. Cherroret, T. Karpiuk, C. A. M\"uller, B. Gr\'emaud, and C. Miniatura,
Phys. Rev. A \textbf{85}, 011604 (2012).

\bibitem{Karpiuk12}
T. Karpiuk, N. Cherroret, K. L. Lee, B. Gr\'emaud, C. A. M\"uller, and C. Miniatura,
Phys. Rev. Lett. \textbf{109}, 190601 (2012).

\bibitem{Micklitz14}
T. Micklitz, C. A. M\"uller, and A. Altland,
Phys. Rev. Lett. \textbf{112}, 110602 (2014).

\bibitem{Loon14}
K. L. Lee, B. Gr\'emaud, and C. Miniatura,
arXiv:1405.2979.

\bibitem{Plisson13}
T. Plisson, T. Bourdel, and C. A. M\"uller,
Eur. J. Phys. ST \textbf{217}, 79 (2013).

\bibitem{Josse12}  
F. Jendrzejewski, K. M\"uller, J. Richard, A. Date, T. Plisson, P. Bouyer, A. Aspect, and V. Josse,
Phys. Rev. Lett. \textbf{109}, 195302 (2012).

\bibitem{Cord14}
T. Micklitz, C. A. M\"uller, and A. Atland,
arXiv:1406.6915.

\bibitem{Aegerter09}
see C. M. Aegerter, G. Maret,
Prog. in Opt. \textbf{52} (2009) and references therein.

\bibitem{ColdDisorderRevs}
D. Cl\'ement, A. F. Var\'on, J. A. Retter, L Sanchez-Palencia, A. Aspect, and P. Bouyer, 
New J. Phys. \textbf{8}, 165 (2006).

\bibitem{Huntley89}
J. M. Huntley, Appl. Opt. \textbf{28}, 4316 (1989).

\bibitem{Horak98}
P. Horak, J.-Y. Courtois, and G. Grynberg, Phys. Rev. A  \textbf{58}, 3953 (1998).

\bibitem{Fehske09}
H. Fehske, J. Schleede, G. Schubert, G. Wellein, V. S. Filinov, and A. R. Bishop,
Phys. Lett. A \textbf{373}, 2182 (2009).

\bibitem{Roche97}
S. Roche and D. Mayou,
Phys. Rev. Lett. \textbf{79}, 2518 (1997).

\bibitem{Kuhn07}
R. C. Kuhn, O. Sigwarth, C. Miniatura, D. Delande, and C. A. M\"uller,
New J. Phys. \textbf{9}, 161 (2007).

\bibitem{AM}
E. Akkermans and G. Montambaux, \emph{Mesoscopic Physics of Electrons and Photons} (Cambridge University Press, 2007).

\bibitem{Shapiro12}
B. Shapiro, J. Phys. A: Math. Theor. \textbf{45}, 143001 (2012).

\bibitem{Thouless74} 
D. J. Thouless,
Phys. Rep. \textbf{13}, 93 (1974).

\bibitem{Efetov99}
K.~B.~Efetov, \emph{Supersymmetry in Disorder and Chaos}, Cambridge University Press, Cambridge, 1997.

\bibitem{Mirlin00}
A.~D.~Mirlin, Phys. Rep. {\bf 326}, 259 (2000).

\bibitem{Prigodin94}
V. N. Prigodin, B. L. Altshuler, K. B. Efetov, and S. Lida, 
Phys. Rev. Lett. \textbf{72}, 546 (1994).

\bibitem{footnote1}
With our definition of $\xi(E)$, the transmission $T(E)$ of a bar of length $L$ is such that $\overline{\ln T(E)}$ scales as $-L/\xi(E)$ at large $L$.

\bibitem{McKinnon83}
A. McKinnon and B. Kramer, 
Z. Phys. B \textbf{53}, 1 (1983).

\bibitem{Slevin14}
K. Slevin and T. Ohtsuki, 
New J. Phys. \textbf{16}, 015012 (2014).

\bibitem{Delande14}
D. Delande and G. Orso,
Phys. Rev. Lett. \textbf{113}, 060601 (2014).


\bibitem{Mott70}
N. F. Mott, Philos. Mag. \textbf{22}, 7 (1970).

\bibitem{Sivan87}
U. Sivan and Y. Imry, Phys. Rev. B \textbf{35}, 6074 (1987).

\bibitem{Atland95}
A. Atland and D. Fuchs, Phys. Rev. Lett. \textbf{74}, 4269 (1995).

\bibitem{Mehta}
M. L. Mehta, \emph{Random matrices}, New York: Academic Press, 1991.

\bibitem{footnote2}
The definition $\tauh\equiv2\pi\overline{\nu}\xi^2$ suffers from an ambiguity in the definition of $\xi$. In particular, we observe that the localization length given by the transfer-matrix approach underestimates the Heisenberg time. We instead evaluate $\tauh$ from the localization length $\tilde{\xi}$ that controls the exponential decay of the average density distribution at large distances. For $V_0=5$ and at energy $E=-2.5$, we find $\tilde{\xi}\simeq 9.8$, which is larger than $\xi$ as expected from the strong wave-function fluctuations in the localization regime \cite{Gogolin88}. A precise estimation of $\tauh$ would require a fine analysis of the spectrum of the system in the localization regime, a task beyond the scope of the present paper.

\bibitem{Gogolin88}
A. A. Gogolin,
Phys. Rep. \textbf{166}, 269 (1988).


\bibitem{Bayer93}
G. Bayer and T. Niederdr\"ank
Phys. Rev. Lett. \textbf{70}, 3884 (1993).

\bibitem{Tourin97}
A. Tourin, A. Derode, P. Roux, B. A. van Tiggelen, and M. Fink,
Phys. Rev. Lett. \textbf{79}, 3637 (1997).

\bibitem{Wolf85}
P.-E. Wolf and G. Maret,
Phys. Rev. Lett. \textbf{55}, 2696 (1985)

\bibitem{Albada85}
M. P. van Albada and A. Lagendijk,
Phys. Rev. Lett. \textbf{55}, 2692 (1985). 

\bibitem{Raedt89}
H. De Raedt, A. Lagendijk, and P. de Vries,
Phys. Rev. Lett. \textbf{62}, 47 (1989).





\bibitem{Cherroret14}
N. Cherroret, B. Vermersch, J. C. Garreau, and D. Delande,
Phys. Rev. Lett. \textbf{112}, 170603 (2014).

\bibitem{Kopidakis08}
G. Kopidakis, S. Komineas, S. Flach, and S. Aubry,
Phys. Rev. Lett. \textbf{100}, 084103 (2008).

\bibitem{Pikovsky08}
A. S. Pikovsky and D. L. Shepelyansky,
Phys. Rev. Lett. \textbf{100}, 094101 (2008).

\bibitem{Basko11}
D. M. Basko,
Ann. Phys. \textbf{326}, 1577 (2011).

\bibitem{Hartung08}
M. Hartung, T. Wellens, C. A. M\"{u}ller, K. Richter, and P. Schlagheck,
Phys. Rev. Lett. \textbf{101}, 020603 (2008).




 
\end{thebibliography}
\end{document}